\documentclass[aps,prb,showpacs,twocolumn,longbibliography]{revtex4-1}
\usepackage[english]{babel}
\usepackage[utf8]{inputenc}
\usepackage{amsmath}
\usepackage{amssymb}
\usepackage[caption=false]{subfig}
\usepackage{amssymb}
\usepackage{epsfig}
\usepackage{graphicx}
\usepackage{amsmath}
\usepackage{array,color}
\usepackage{natbib}

\usepackage[usenames,dvipsnames]{xcolor}
\definecolor{forestgreen}{rgb}{0.11,0.54,0.15}
\definecolor{purple}{rgb}{0.62,0.10,0.96}
\definecolor{dockerblue}{rgb}{0.11,0.56,0.98}
\definecolor{freeblue}{rgb}{0.25,0.41,0.88}

\usepackage[pdftex,plainpages=false,colorlinks=true,linkcolor=Red, citecolor=blue, urlcolor=blue]{hyperref}

\newcommand{\Tr}{\text{Tr}}

\newcommand{\eref}[1]{Eq.~(\ref{#1})}

\begin{document}
\title{Orbital magnetization and anomalous Hall effect in interacting Weyl semimetals}
\author{S. Acheche$^{1}$, R. Nourafkan$^{1}$ and A.-M. S. Tremblay$^{1,2}$}
\affiliation{
$^1$D\'{e}partement de physique, Institut quantique, and Regroupement qu\'eb\'ecois sur les mat\'eriaux de pointe, Universit\'{e} de Sherbrooke, Sherbrooke, Qu\'{e}bec, Canada J1K 2R1 \\
$^2$Canadian Institute for Advanced Research, Toronto, Ontario, Canada, M5G 1Z8\\
}
\date{\today}
\begin{abstract}
Ferromagnetic Weyl semi-metals exhibit an anomalous Hall effect, a consequence of their topological properties. In the non-interacting case, the derivative of the orbital magnetization with respect to chemical potential is proportional to this anomalous Hall effect, the St\v{r}eda formula. Motivated by compounds such as $\text{Mn}_3\text{Sn}$, here we investigate how interactions modeled by a Hubbard $U$ impact on both quantities when the Fermi energy is either aligned with the Weyl nodes or away from them. Using Dynamical Mean-Field Theory, we first find interaction-induced Mott- and band-insulating transitions. In the Weyl semimetal regime, away from insulators, interactions lead to an increase in the imbalance between the densities of spin species induced by a Zeeman term $h$. This increased imbalance leads to an increase of the anomalous Hall effect that can also be understood from the displacement of the Weyl nodes and topological arguments. This interaction-induced spin imbalance also compensates the reduction in orbital magnetization of each spin species that comes from smaller quasiparticle weight. In the small inteaction regime, the combined effects lead to an orbital magnetization that depends weakly on interaction and still changes linearly with chemical potential at small doping. In the intermediate and strong correlation regimes, the localization due to interaction affects strongly the orbital magnetization, which becomes small. A quasiparticle picture explains the anomalous Hall effect but does not suffice for the orbital magnetization. We propose a modified St\v{r}eda formula to relate anomalous Hall effect and orbital magnetization in the weak correlation limit. We also identify mirror and particle-hole symmetries of the lattice model that explain, respectively, the vanishing of the anomalous Hall effect at $h=0$ for all $U$ and of the orbital magnetization at half-filling, $\mu=0$, for all $U$ and $h$.

\end{abstract}
\maketitle



\section{Introduction}

Weyl semimetals are three-dimensional (3D) topological materials in which two non-degenerate energy bands touch at isolated wave-vectors (Weyl nodes)~\cite{Armitage_2018, Yan_2017}. The electronic structure near the Weyl nodes is described by linearly dispersive bands. The  nondegenerate bands are a consequence of either time-reversal or inversion symmetry breaking. The Weyl physics governs the transport properties when the Fermi energy lies on, or in the vicinity, of the Weyl nodes. For instance, a subclass of Weyl semimetals named ferromagnetic (FM) Weyl metals demonstrates a non-zero anomalous Hall effect (AHE)~\cite{Burkov_2014,Ahn_2017}.

The AHE is not a specific property of FM Weyl metals since every FM conductor can potentially exhibit a non-zero transverse conductivity which is not related to an external magnetic field~\cite{Nagaosa_2010}. Two main contributions to the Hall conductivity are (i) an intrinsic AHE that comes from Berry curvature and (ii) an extrinsic AHE due to scattering on impurities in the presence of spin-orbit coupling~\cite{Nagaosa_2010}. In general, both contributions are present with the same order of magnitude, but for FM Weyl metals the intrinsic AHE tends to dominate the extrinsic part even if the chemical potential does not lie exactly on the Weyl nodes\cite{Burkov_2014}. This theoretical assertion has found a possible experimental evidence recently with the Weyl ferromagnet Co$_3$Sn$_2$S$_2$~\cite{Wang_2018}.

A non-zero Berry curvature not only influences transport properties but also impacts properties such as orbital magnetization~\cite{Xiao_2005, Thonhauser_2005, Nourafkan_2014,Aryasetiawan_2016, Shi_2007,Ogata_2015,Ogata_2016}.
Indeed, in a non-interacting system the St\v{r}eda formula~\cite{Streda_1982} links the orbital magnetization and intrinsic AHE, the latter being proportional to the derivative of the magnetization with respect to the chemical potential. In the case of FM Weyl metals, the St\v{r}eda formula suggests the common origin of the two quantities: Both owe their existence to stacks of two-dimensional Chern insulators in momentum space~\cite{Hosur_2013,Thonhauser_2005}.

How electronic correlations impact the AHE and orbital magnetization of Weyl semimetals is the subject of this study. In particular, recent experimental evidence has shown  Weyl physics in the presence of bands that are strongly renormalized by interactions~\cite{Kuroda_2017}. For instance, the Weyl material $\text{Mn}_3\text{Sn}$\cite{Kuroda_2017}, an antiferromagnet with weak ferromagnetism, is known to exhibit a large AHE even at room temperature\cite{Nakatsuji_2015}. All this motivates our study of the interplay between electron-electron interactions and topological properties of Weyl semimetals. 

Theoretically, interacting Weyl semimetals have been studied through various methods, such as random phase approximation \cite{Hofmann_2015}, renormalization group \cite{Maciejko_2014,Roy_2017} and Hubbard interaction through cluster methods\cite{Krempa_2014,Laubach_2016}. The focus of attention was broken-symmetry phases, band renormalization or stability of Weyl nodes under the effect of electron-electron interaction. In this paper, we investigate how interactions represented by a Hubbard $U$ on a lattice model of Weyl semimetals impact on the AHE and on orbital magnetization of the FM Weyl semi-metal. We show how the St\v{r}eda formula is changed due to the presence of interaction and how many of the results can be explained from a quasiparticle perspective. We also find that because of competing effects, the orbital magnetization is rather insensitive to interactions near half-filling, even though it is not a purely topological quantity. We briefly discuss also the semi-metal to insulator transition at large $U$. 

The model and method are presented in section \ref{Sec:Model_method}. Section \ref{Sec:Interaction_WSM} introduces various properties of interacting Weyl semimetals, in particular how the metal-insulator transition occurs either through Mott physics or through Weyl-node collapse. In section \ref{Sec:AHE}, we compute the Hall conductivity to see how it changes as a function of interaction. A quasiparticle (QP) approach is derived in order to capture this change in a single-particle formalism. Finally, Section \ref{Sec:Orbital_magnetization} is devoted to the orbital magnetization. We check the validity of the St\v{r}eda formula for our model in the non-interacting case and suggest how to modify this formula in the presence of the Hubbard interaction. The QP approach and perturbation theory are used there again to understand quantitatively the interacting case.

\section{Model and method}\label{Sec:Model_method}

We use a lattice model of Weyl semimetals with Hubbard interaction on a cubic lattice. The model is simple and is usually considered as a generic model for time-reversal symmetry breaking Weyl semimetals~\cite{Roy_2016}. We will see however that it has some symmetries that are not obvious at first sight. The full Hamiltonian is
\begin{equation}\label{EqHamiltonian_total}
\hat{H} = \hat{H}_0 + U\sum_{\textbf{r}} \left(\hat{n}_{\textbf{r}\uparrow}-\frac{1}{2}\right) \left(\hat{n}_{\textbf{r}\downarrow}-\frac{1}{2}\right) - \mu \sum_{\textbf{r}} \hat{n}_{\textbf{r}}, 
\end{equation}
where $U$ is the Hubbard interaction term, $\mu$ is the chemical potential and $\hat{H}_0$ denotes the non-interacting part of the Hamiltonian. Its momentum-space expression is~\cite{Roy_2016}
\begin{eqnarray} \label{Eq.Hamiltonian_nonint}
\nonumber
\hat{H}_0 = &\sum_{\textbf{k}}& \hat{\mathbf{C}}^{\dagger}_{\textbf{k}} \left[\left\lbrace h - 2t\left(\cos k_x + \cos k_y + \cos k_z\right) \right\rbrace \boldsymbol{\sigma}_z \right. \\
&+& \left. 2t \sin k_y \boldsymbol{\sigma}_y+ 2t \sin k_x \boldsymbol{\sigma}_x \right] \hat{\mathbf{C}}_{\textbf{k}}.
\end{eqnarray}
Here $t$ denotes the nearest-neighbor hopping amplitude and $h$ is a Zeeman term. We set $t = 1$ as energy unit and use Hartree units ($\hbar = e = k_B = 1$). The destruction (creation) operator $\hat{\mathbf{C}}_\mathbf{k} = (\hat{c}_{\mathbf{k}\uparrow} , \hat{c}_{\mathbf{k}\downarrow})^{(T)}$ is represented in the spin basis and $\boldsymbol{\sigma}_i$ are Pauli matrices.
For those values of $h$ used in this paper, there are four Weyl nodes whose location and chirality are indicated in Table~I
\begin{center}
\begin{table}\label{Table:WeylLocation}
\begin{tabular}
{|l|l|l|}\hline
Chirality &  & \\\hline
$-1$ & $\left(  \pi,0,+\arccos\left(  h/2t\right)  \right)  $ & $\left(
0,\pi,+\arccos\left(  h/2t\right)  \right)  $\\\hline
$+1$ & $\left(  \pi,0,-\arccos\left(  h/2t\right)  \right)  $ & $\left(
0,\pi,-\arccos\left(  h/2t\right)  \right)  $\\\hline
\end{tabular}
\caption{Location and chirality of the 4 Weyl nodes for $|h|<2t$.}
\end{table}
\end{center}
The Hamiltonian has several symmetries that will help us later to understand the physics of the results. First, let us define elementary discrete transformations that have eigenvalues $\pm 1$: Parity $P$, time reversal $T$, translation $\mathcal{T}_\mathbf{Q}$ by $\mathbf{Q}=(\pi,\pi,\pi)$ , $xz$ mirror $\mathcal{M}_{xz}$, and $yz$ mirror $\mathcal{M}_{yz}$. Note that $\mathcal{T}_\mathbf{Q}$ reflects the bipartite nature of the lattice. We also define a pseudo charge conjugation transformation $C'$ analogous to, but different from, the one that usually appears in field theory~\cite{Weinberg}. The effect of these elementary transformations on the destruction operators and their eigenvalues for chirality and spin components of the non-interacting Hamiltonian Eq.~\ref{Eq.Hamiltonian_nonint} in the case $h=0$ are tabulated in Table~\ref{Table:DiscreteSymmetries}. The transformations act on the same way on the labels of the creation operators.  

\begin{center}%
\begin{table}     \label{Table:DiscreteSymmetries}
    \begin{tabular}
    {|c|c|c|c|c|}\hline
    Discrete transformations & $\chi$ & ${\sigma}_{x}$ & ${\sigma}_{y}$ & ${\sigma}_{z}$ \\\hline
    $P\hat{c}_{\mathbf{k\sigma}}P^{-1}= \hat{c}_{-\mathbf{k\sigma}}$ & $-1$ & $+1$ &
    $+1$ & $+1$\\\hline
    $T\hat{c}_\mathbf{k\sigma}T^{-1}=i\sigma^y_{\sigma\sigma'} \hat{c}_{-\mathbf{k-\sigma'}}$ & $-1$ &
    $-1$ & $-1$ & $-1$\\\hline
    $C^{\prime}\hat{c}_{\mathbf{k\sigma}}C^{\prime-1}= \hat{c}_{\mathbf{k\sigma}%
    }^{\dagger}$ & $+1$ & $-1$ & $-1$ & $-1$\\\hline
    $\mathcal{T}_{\mathbf{Q}}\hat{c}_{\mathbf{k\sigma}}\mathcal{T}_{\mathbf{Q}}^{-1}%
    = \hat{c}_{\mathbf{k+Q\sigma}}$ & $-1$ & $+1$ & $+1$ & $+1$ \\\hline
    $\mathcal{M}_{zy}\hat{c}_{\mathbf{k\sigma}}\mathcal{M}_{zy}^{-1}= \sigma^x_{\sigma\sigma'}\hat{c}_{(-k_{x}%
    ,k_{y,}k_{z})\mathbf{\sigma'}}$ & $-1$ & $+1$ & $-1$ & $-1$ \\\hline
    $\mathcal{M}_{zx}\hat{c}_{\mathbf{k\sigma}}\mathcal{M}_{zx}^{-1}= i\sigma^y_{\sigma\sigma'} \hat{c}_{(k_{x},-k_{y,}k_{z})\mathbf{\sigma'}}$ & $-1$ & $-1$ & $+1$ & $-1$\\\hline
    \end{tabular}

    \caption{Effect of elementary transformations on the destruction operators along with their effect on $\chi$ and on the the spin eigenstates of $\sigma_{x,y,z}$. 
    $P$ is parity, $T$ time reversal, $C'$ pseudo charge conjugation, $\mathcal{T}_\mathcal{Q}$ translation by $\mathcal{Q}=(\pi,\pi,\pi)$, and $\mathcal{M}_{zx\ /\ zy}$ mirrors across the specified plane. Summation over repeated $\sigma'$ is implied.}
    \end{table}
    \end{center}

Using these results, we find that when $\mu=0$, the following symmetry operation,     
\begin{equation}\label{Eq:particle_hole}
(C'PT) \hat{c}_{\mathbf{k\sigma}}(C'PT)^{-1}=i\sigma^y_{\sigma\sigma'} \hat{c}^\dagger_{\mathbf{k\sigma'}},
\end{equation}   
is a particle-hole transformation that leaves Eq.~\ref{EqHamiltonian_total} invariant for all $U$, and $h$. Since under this symmetry operation $\hat{n}_{\sigma}\rightarrow 1-\hat{n}_{-\sigma}$, this means that $\mu=0$ corresponds to half-filling. 

When $h=0$ , but for arbitrary values of $\mu$ and $U$, there are two more symmetries of $\hat{H}$ that we will call \textit{translated mirrors}, for lack of a better name
\begin{eqnarray}
(\mathcal{T}_{\mathbf{Q}}\mathcal{M}_{zy})\hat{c}_{\mathbf{k\sigma}}(\mathcal{T}_{\mathbf{Q}}\mathcal{M}_{zy})^{-1}&=& \sigma^x_{\sigma\sigma'}\hat{c}_{(-k_{x}+\pi,k_{y}+\pi,k_{z}+\pi)\mathbf{\sigma'}} \nonumber \\ \label{Eq.symA} \\
(\mathcal{T}_{\mathbf{Q}}\mathcal{M}_{zx})\hat{c}_{\mathbf{k\sigma}}(\mathcal{T}_{\mathbf{Q}}\mathcal{M}_{zx})^{-1}&=& i\sigma^y_{\sigma\sigma'}\hat{c}_{(k_{x}+\pi,-k_{y}+\pi,k_{z}+\pi)\mathbf{\sigma'}} \nonumber \\ 
\label{Eq.symB}
\end{eqnarray}

In this paper, the fully interacting system is solved within the dynamical mean-field theory (DMFT) framework\cite{Georges_1996}. In this theory, the system is mapped onto an interacting quantum impurity site hybridized with a bath of non-interacting electrons. The lattice self-energy is the same as the impurity self-energy. The hybridization function between the impurity and the bath is then determined self-consistently by imposing equality of the impurity Green's function and the projection of the lattice Green's function on a single site. Although the method is exact only in the limit of infinite dimension, it has been very sucessful in describing the physics of three-dimensional systems. To obtain the impurity Green's function, a finite temperature exact diagonalization (ED) solver~\cite{Caffarel_1994} is used. For an ED solver, the non-interacting electron bath is represented by a finite number of bath \textit{sites} hybridized with the impurity. The impurity problem is then solved exactly by diagonalizing numerically the Hamiltonian. In ED, the self-consistency condition is partially respected by minimizing with respect to the bath parameters the mean-square distance between the hybridization function obtained from Dyson equation and the one written in terms of bath parameters~\cite{Caffarel_1994}. We usually keep 1024 Matsubara frequencies. The convergence criterion is $2\times 10^{-5}$ on a relative scale. All numerical computations are done at inverse temperature $\beta=80$ and with five bath sites. Increasing the number of bath sites should lead to more quantitatively accurate results but should not change our conclusions. The main effect of the finite number of bath sites is that real-frequency quantities are represented by a set of closely spaced delta functions. We will see a consequence of that in the next section. The spacing increases at larger frequencies. However, quantities in Matsubara frequencies, which is all that we need for thermodynamic quantities such as orbital magnetization and AHE, remain smooth functions that are accurately represented by the finite number of bath sites.  

\section{Interacting Weyl semimetal and Mott transition}\label{Sec:Interaction_WSM}

\begin{figure}
\begin{center}
\includegraphics[scale=0.38]{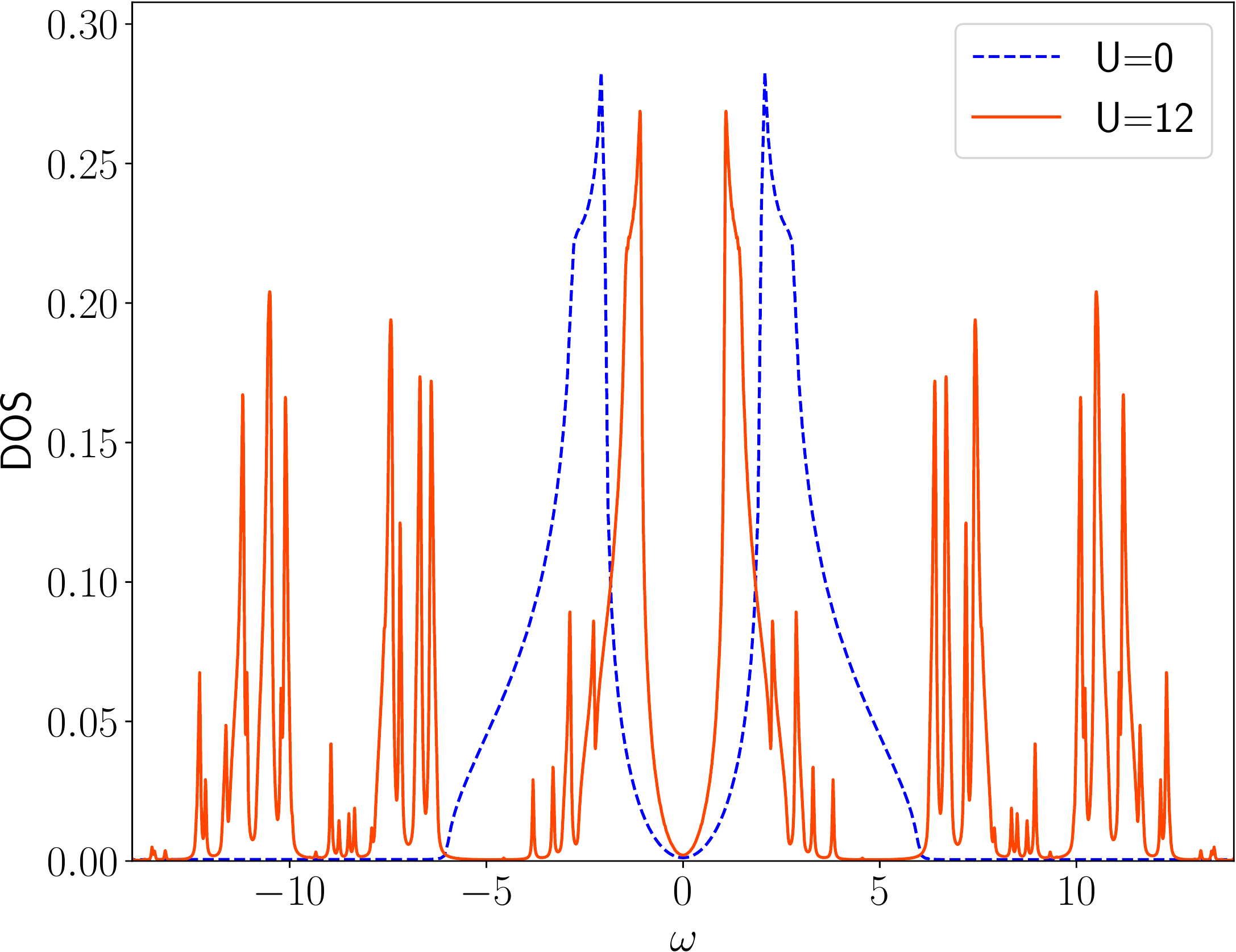}
\caption{(Color online) (a) Density of states for non interacting (dashed blue line) and interacting (continuous red line) Weyl semimetal. $U=12$ corresponds to the bandwidth of the model \eref{Eq.Hamiltonian_nonint}. Sharp peaks are due to the finite number of bath sites of the DMFT impurity solver. We use a Lorentzian broadening $\eta = 0.02$ for both densities.}\label{Fig.DOS}
\end{center}
\end{figure}

In this section we discuss how interactions impact the dynamics of the Weyl electrons, first in the absence of a Zeeman coupling ($h=0$). At a finite value of $U$, several of the changes occuring in the system manifest themselves in the density of states (DOS). In figure~\ref{Fig.DOS}, we show the DOS for both non-interacting and interacting Weyl semimetals. In the non-interacting case, two regions can be defined: Near the Fermi level \emph{i.e.} for $-2 \lesssim \omega \lesssim 2$, electronic bands are mainly linear and we recover the DOS of three-dimensional (3D) semimetals characterized by $\omega^2$ behavior. Outside this region, the dispersion is not linear anymore and the DOS eventually retains the characteristics of the cubic lattice DOS. In the presence of the Hubbard interaction, band energies are renormalized leading to a narrower DOS near the Fermi level and to Hubbard band satellites. The renormalization factor is given by the bulk quasiparticle weight defined as $\boldsymbol{Z} \simeq \left[\textbf{I} - \text{Im} \boldsymbol{\Sigma}(i\omega_0)/\omega_0\right]^{-1}$, where $\boldsymbol{\Sigma}$ denotes the Matsubara self-energy and $\omega_0 = \pi/\beta$ is the first fermionic Matsubara frequency. In our system at half-filling, $\boldsymbol{Z}$ is a diagonal matrix with $Z_{\uparrow \uparrow} =Z_{\downarrow\downarrow}=Z$.  Hence, the low-energy effective Hamiltonian describing the Weyl fermion physics is obtained from the non-interacting one by subtituting for the velocity tensor $v_{ij} \rightarrow Z v_{ij}$. Note that the incoherent part (Hubbard bands) does not contribute to transport properties, so a renomalized non-interacting low-energy Hamiltonian should be a good approximation for determining these properties.

As expected, the lower and upper Hubbard bands form respectively at $\omega = - U/2$ and $\omega = U/2$ by transferring spectral weight to high energy. At $U_{c2} \approx 19.5$, the system undergoes a phase transition from a semimetal to a Mott insulating phase. The Mott phase has trivial topology with a large charge gap and is characterized by the emergence of a diverging self-energy at low Matsubara frequencies. The Mott transition occurs when the Hubbard interaction is large enough to localize electrons on lattice sites. It is worth mentioning that the interaction does not shift Weyl nodes when $h=0$ even away from half-filling.
This property is not related to the locality of the DMFT self-energy but rather symmetries of the microscopic model, as we discuss below. 
In other lattice models, \emph{e.g.} in Ref.~\onlinecite{Krempa_2014} and~\onlinecite{Laubach_2016}, Weyl-node positions depend on the prefactor of $\boldsymbol{\sigma}_x$. In this case, the Fock diagram shifts Weyl nodes in the same way as the Hartree diagram shifts nodes in our case when $h$ is finite. 
Indeed, in the finite $h$ case, the Weyl-node locations in our model depend on the prefactor $h$ of $\boldsymbol{\sigma}_z$ in the non-interacting Hamiltonian. A finite value of $h$ leads to a self-energy proportional to $\boldsymbol{\sigma}_z$ that changes this prefactor. 

We can understand the above results with the help of symmetries. In the absence of a Zeeman term ($h=0$), the translated mirror symmetries~\eref{Eq.symA} and~\eref{Eq.symB} fix the position of Weyl nodes in the Brillouin zone. Since these symmetries hold independently of the electronic density of the system, the Weyl nodes are fixed also away from half-filling. A finite Zeeman term $h$ breaks those symmetries and makes the postion of Weyl nodes interaction dependent.

The shift in position of Weyl nodes in the presence of a finite $h$ leads to two different scenarios for the insulating phases in the strong correlation regime~\cite{Zhu_2017}: on one hand, if $U$ becomes large enough before merging of the Weyl nodes, the system goes to the Mott insulating phase as in the case $h=0$. On the other hand, if Weyl nodes merge before the Mott transition, the system becomes a \textit{band insulator} characterized by full spin-polarization. Since only one species of spin is present once this semimetal to band-insulator transition occurs, the Hubbard interaction has no effect anymore, regardless of the value of $U$.

\section{Anomalous Hall effect}\label{Sec:AHE}

The anomalous Hall conductivity $\sigma_{xy}$ is a dissipation-less transverse current response to an electric field. It is given by the first order expansion in powers of frequency of the corresponding current-current correlation function,  $\Pi_{xy}$, in frequency. In contrast to an Ohmic (dissipative) response, the frequency expansion can be done on the imaginary axis.~\cite{Nourafkan_2013} Using the standard formula for the polarization $\Pi_{xy}$,~\cite{Mahan} we have
\begin{widetext}
\begin{eqnarray}\label{AHC}
\Pi_{xy}(i\nu_n) &=& -\frac{1}{N\beta} \sum_{\textbf{k}\omega_m} \Tr\left[\boldsymbol{v}_x(\textbf{k})\boldsymbol{G}(i\omega_m+i\nu_n,\textbf{k})\frac{\partial \boldsymbol{G}^{-1}(i\omega_m,\textbf{k})}{\partial k_y}\boldsymbol{G}(i\omega_m, \textbf{k}) \right] \\
\nonumber
 &\approx&   \Pi_{xy}(0)+ \underbrace{\frac{1}{N\beta} \sum_{\textbf{k}\omega_m} \Tr\left[ 
\frac{\partial \boldsymbol{G}^{-1}(i\omega_m,\textbf{k})}{\partial k_x}
\boldsymbol{G}(i\omega_m,\textbf{k})\frac{\partial \boldsymbol{G}^{-1}(i\omega_m,\textbf{k})}{\partial i\omega_n}\boldsymbol{G}(i\omega_m, \textbf{k})\frac{\partial \boldsymbol{G}^{-1}(i\omega_m,\textbf{k})}{\partial k_y}\boldsymbol{G}(i\omega_m, \textbf{k}) \right]}_{\text{first order}}i\nu_n +\dots 
\end{eqnarray}
\end{widetext}
where $\nu_n$ denotes bosonic Matsubara frequencies. The velocity operators are defined as $\boldsymbol{v}_i = -\partial_{k_i} \boldsymbol{H}_0$ and $\boldsymbol{G}$ is the interacting Green's function given by
\begin{equation}
\boldsymbol{G}(i\omega_n, \textbf{k}) = \left[(i\omega_n+\mu)\textbf{I} - \boldsymbol{H}_0(\textbf{k}) - \boldsymbol{\Sigma}(i\omega_n) \right]^{-1}.
\end{equation}
In DMFT, the self-energy has only frequency dependence, hence the derivatives of inverse Green's functions with respect to wave vectors reduce to velocity operators without vertex corrections. However, the derivative of the self-energy with respect to frequency is taken into account. 
%

\begin{figure}
\begin{center}
\includegraphics[scale=0.7]{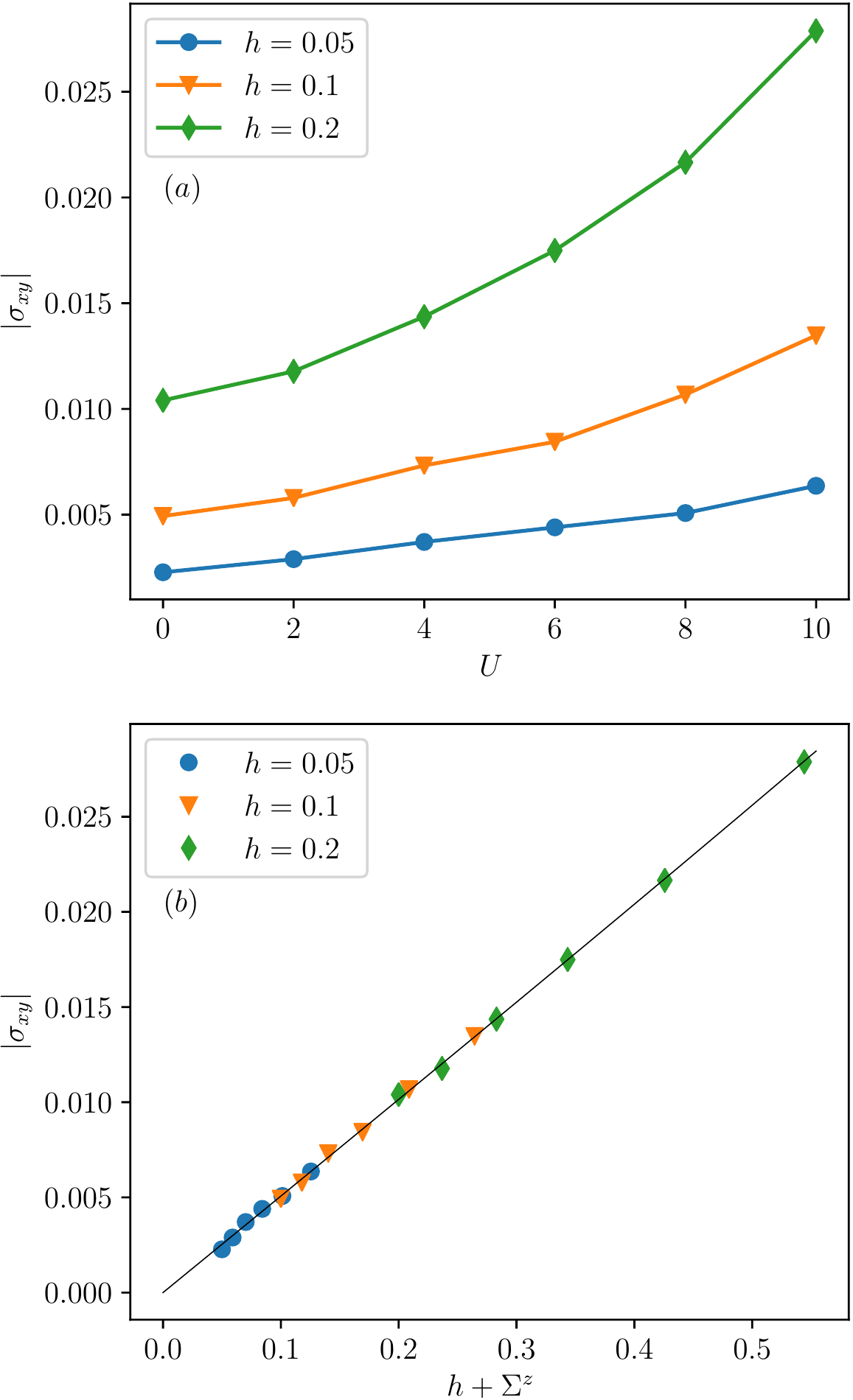}
\caption{(a) Anomalous Hall conductivity as a function of $U$ at half-filling. (b) Anomalous Hall conductivity as a function of the renormalized Zeeman term $\tilde{h}= h +\Sigma^z$. The continous black line is the theoretical value found by replacing $h$ by $\tilde{h}$ in \eref{Eq.value_sigdc}.}\label{Fig.sigmaxy}
\end{center}
\end{figure}
%

The non-trivial band topology of a FM Weyl semimetal gives a finite AHE which can be understood as follows; in reciprocal space, each Weyl node can be seen as the position of a \textit{magnetic} monopole, where the associated magnetic field is the Berry curvature.~\cite{Xiao_2010} Assuming only two Weyl nodes with opposite chirality separated by a vector $\textbf{b}$ in the first Brillouin zone, the region between nodes can be decomposed as a stack of two-dimensional (2D) Chern insulators. Each of these insulators carries a non-zero Chern number $C$ corresponding to the flux of the Berry curvature in a plane and contributes to the anomalous conductivity~\cite{Burkov_2011,Yang_2011, Xu_2011, Hosur_2013, Haldane_2004}. For example, if $\textbf{b}$ is along the $z$ direction, then the AHE in the $z$ direction is given by
\begin{equation}\label{Eq.sigmaxyTopo}
\sigma_{xy} =\frac{C b}{2\pi^2},
\end{equation}
which depends only on the Weyl-node separation in wave-vector space. 
In systems with several Weyl nodes, one can define an effective vector $\textbf{b}_{eff}= \sum_i \chi_i \textbf{b}_i$, where the sum is over all Weyl nodes in the first Brillouin zone and $\chi_i$ is the chirality of the $i$~the Weyl node located at $\textbf{b}_i$  with respect to an arbitrary origin\footnote{One has to be careful not substitute blindly $\textbf{b}_{eff}$ in equation \ref{Eq.sigmaxyTopo}. The periodicity of the Brillouin zone in lattice models needs to be taken into account. This explains the absence of the AHE in our case when $h=0$.}. 

As presented in Table~\ref{Table:WeylLocation}, \eref{Eq.Hamiltonian_nonint} has four Weyl nodes in the Brillouin zone located at $(0, \pi, \pm \arccos(h/2t))$ and $(\pi,0, \pm \arccos(h/2t))$. 
In the absence of Zeeman coupling ($h=0$), the $\textbf{b}_{eff}$ vector is $(0 ,0 ,2\pi)$, which is equivalent to a vanishing effective vector, yielding  zero anomalous Hall conductivity.
Moreover, one can show by contradiction that at $h=0$ the translated mirror symmetries \eref{Eq.symA} and \eref{Eq.symB} fix the position of the Weyl nodes to $(0, \pi, \pm \pi/2)$ and $(\pi,0, \pm \pi/2)$ even when interactions $U$ are present, hence the AHE vanishes also in this case.
 
Alternatively, using an effective Hamiltonian where lattice effects are kept only in the $z$ direction, (See appendix~\ref{Sec:AppendixA}), when $h=0$ the intrinsic part of anomalous Hall conductivity is given by (assuming the chemical potential lies on the node)
\begin{equation}\label{Eq2}
\tilde{\sigma}_{xy} = -\int dk_z d^2q\frac{2t^2(-2t\cos(k_z))}{\left((2t\cos(k_z)))^2 +4t^2q^2_x +4t^2q^2_y\right)^{\frac{3}{2}}},
\end{equation}
where $q_x$ and $q_y$ denote wave vectors with respect to the Weyl node location in the Brillouin zone. As one can see, integrating on $k_z$ leads to a vanishing AHE.\\

A Zeeman term $h\neq 0$ creates a spin-polarized FM Weyl metal, \emph{i.e} a Weyl semimetal where the densities of spin species are not equal to each other ($ \langle n_\uparrow  \rangle \neq \langle n_\downarrow \rangle $). In addition to that, $h$ changes the position of the Weyl nodes in the Brillouin zone, leading also to a non-zero $\textbf{b}_{eff}$ and a finite AHE.

In the presence of a finite $h$, let us consider the non-interacting and interacting cases in turn. At $U=0$, one can use the same effective Hamiltonian as in \eref{Eq2} to compute the anomalous Hall conductivity and find (cf. Appendix~\ref{Sec:AppendixA}):
\begin{equation}\label{Eq.value_sigdc}
\sigma_{xy} = (-1/\pi^2)\arcsin(\frac{h}{2t}).
\end{equation}
Note however that the spin-density imbalance is also the key to understand AHE in FM coumpounds and thus the AHE cannot be attributed to a specific property of Weyl metals.

At finite $U$ we find an effect on the AHE even at half-filling. Panel (a) in figure~\ref{Fig.sigmaxy} shows the change of the Hall conductivity as a function of Hubbard interaction. One recovers the value given by~\eref{Eq.value_sigdc} when $U=0$.  Upon increasing $U>0$, the (absolute value of) the Hall conductivity increases monotonously~\cite{Krempa_2014} despite a reduction in coherent spectral weight. Indeed, due to its particular form,  the Hubbard interaction increases the difference between electronic densities of each spin species leading to an enhanced effective $h$ and, consequently, to a larger value of the AHE. Increasing $U$ for a fixed $h$, the anomalous Hall conductivity keeps increasing until the critical $U$ for the Mott transition where it falls to zero discontinuously. 

 The change in the anomalous Hall conductivity can be quantitatively understood within the QP approximation (see Appendix~\ref{Sec:AppendixB} for more details). With this approximation, the QP Hamiltonian is:
\begin{equation}
\boldsymbol{H}^{QP} = \boldsymbol{Z}^{1/2}\left[\boldsymbol{H}_0 -\mu \textbf{I} + \text{Re} \boldsymbol{\Sigma} (\omega = 0) \right]\boldsymbol{Z}^{1/2},
\end{equation} 
where $\boldsymbol{Z}$ is  the quasiparticle weight.  In the presence of $h$, interaction changes the location of the Weyl nodes to $(0, \pi, \pm \arccos((h+\Sigma^z)/2t))$ and $(\pi,0, \pm \arccos((h+\Sigma^z)/2t))$ leading to a $\textbf{b}_{eff}$ vector that depends on $U$. Figure~\ref{Fig.sigmaxy} shows the change of the AHE as a function of $\tilde{h}=h+\Sigma^z$ for different value of $h$: The black continuous line is the theoretical value presented in~\eref{Eq.value_sigdc} with $h$ replaced by $\tilde{h}$. There is a fine agreement between the analytic expression and numerical computation for a wide range of $U$. 

To understand the expression for $\tilde{h}$, the self-energy at zero frequency has then been rewritten as $\text{Re} \boldsymbol{\Sigma} (\omega=0) = \Sigma^{\text{I}}\textbf{I} + \Sigma^{z} \boldsymbol{\sigma}^z$.  Here, $\Sigma^{\text{I}}=(\Sigma_{\uparrow \uparrow}+\Sigma_{\downarrow \downarrow})/2$  and $\Sigma^z=(\Sigma_{\uparrow \uparrow}-\Sigma_{\downarrow \downarrow})/2$. We performed a perturbative calculation and found that at low to intermediate interaction strengths, $\Sigma^{\text{I}}$ is well described by second-order perturbation theory. However, describing  $\Sigma^{z}$ requires higher-order perturbation theory.

Away from half-filling, there are contributions to the anomalous Hall effect that are not purely topological even in the non-interacting case, so we do not consider the $n\ne 1$ case. 

\section{Orbital magnetization} \label{Sec:Orbital_magnetization}
The impact of non-trivial topology on orbital magnetic susceptibility has been intensively studied\cite{Raoux_2014,Piechon_2016,Rubio_2016, Raoux_2015, Ogata_2016b,Gao_2015,Shitade_2018}. Here, we focus on orbital magnetization, a quantity closely related to the AHE. 
The orbital magnetization is defined by $M^{a}_{\text{orb}}=-\partial (E-\mu N)/ \partial B |_{T=0,B=0}$, where $E$ is the energy of the system. In general, it can be written in terms of the fully interacting Green's function~\cite{Nourafkan_2014} and is approximately given by
\begin{widetext}
\begin{equation}\label{Eq.orbital_diagram}
M^{a}_{\text{orb}}  \simeq  \left( \frac{i}{2N\beta} \right) \sum_{\textbf{k},i\omega_m} \epsilon^{abc}\Tr \left\lbrace  \underbrace{\left[ \boldsymbol{H}_0- \mu \textbf{I} + \frac{\boldsymbol{\Sigma}}{2} \right]}_{\text{energy vertex}} \boldsymbol{G} \left(-\frac{\partial \boldsymbol{G}^{-1}}{\partial k_b} \right) \boldsymbol{G} \left(-\frac{\partial \boldsymbol{G}^{-1}}{\partial k_c} \right) \boldsymbol{G}\right\rbrace.
\end{equation}
\end{widetext}
In ~\eref{Eq.orbital_diagram},  $\epsilon^{abc}$ is the fully anti-symmetric Levi-Civita tensor and $a$,$b$ and $c$ are spatial coordinates ($\{x,y,z \}$). In the paradigm of a purely local self-energy, this formula takes into account both kinetic and potential energy contributions correctly. Indeed, we neglected an additive term proportional to the linear response of the self-energy to the magnetic field but this term vahishes in the DMFT approximation.\\

It is important to stress that this formula reduces to \emph{the modern theory of the orbital magnetization}~\cite{Xiao_2005, Thonhauser_2005} when the self-energy is set to zero (cf. Appendix A of Ref.~\onlinecite{Nourafkan_2014}). This fact indicates that the above general fomula takes into account topological contributions even in the presence of correlations. The formula is derived in the limit of zero external magnetic field. However, we use it in the presence of a small Zeeman term entering $H_0$. This should be correct if the Zeeman term is sufficiently small or if it arises from spontaneous symmetry breaking. We also neglect the spin contribution to the magnetic field and focus on the $z$ component of the orbital magnetization. 

We consider in turn the non-interacting case, where we recover the St\v{r}eda formula that relates anomalous Hall effect and orbital magnetization, and the interacting case, where strong deviations are found. 

\begin{figure*}[t]
\begin{center}
\includegraphics[scale=0.38]{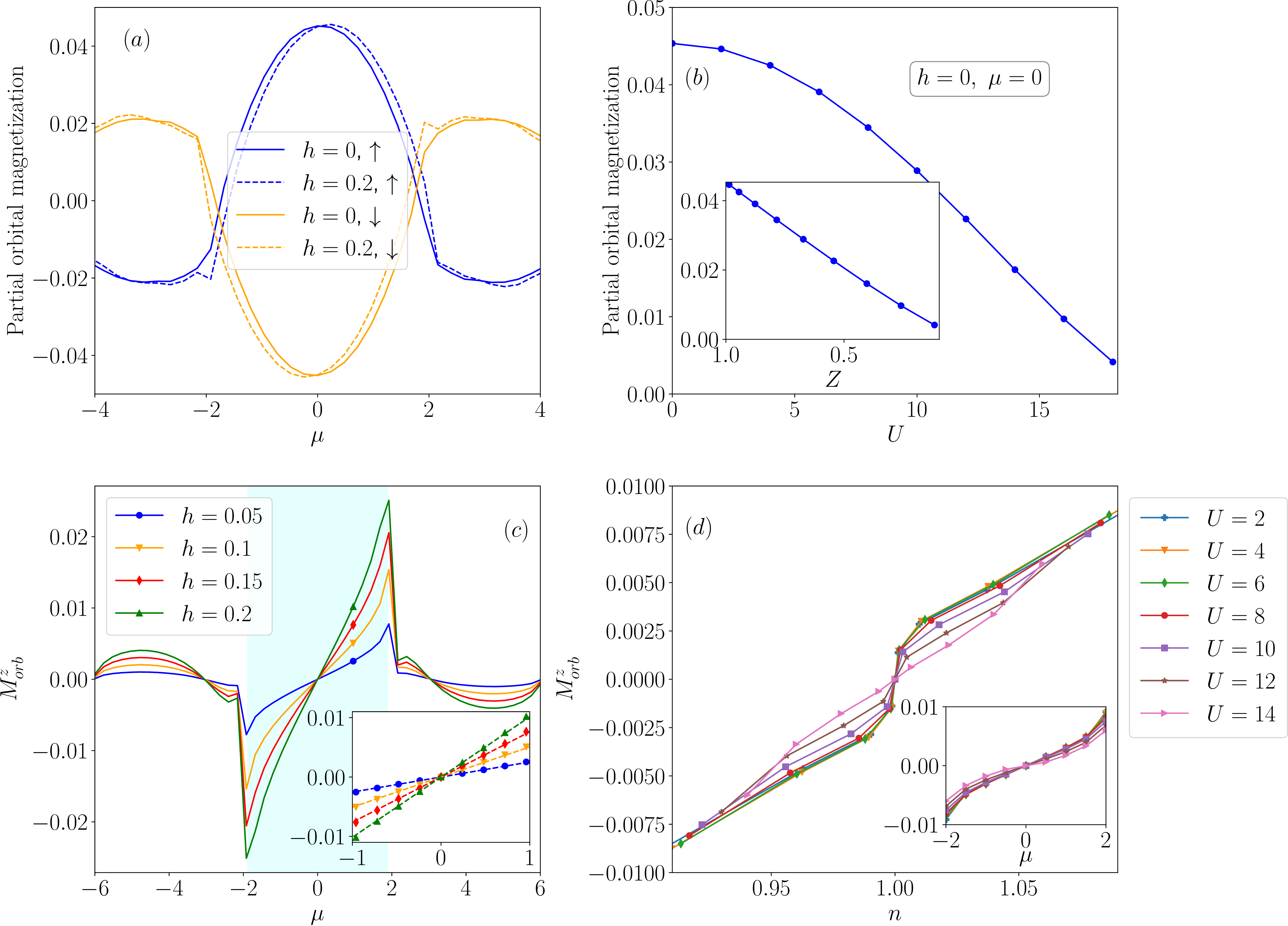}
\caption{(a) Non-interacting orbital partial magnetization for $h=0$ (continuous lines) and $h=0.2$ (dashed lines) as a function of the chemical potential $\mu$. Note that the partial orbital magnetization is basis dependent and we have choosed the spin basis to represent it. (b) Value of the positive partial orbital magnetization as a function of Hubbard interaction $U$ for $h=0$ and $\mu=0$. Interaction renormalizes orbital magnetization via $Z$ as shown in the inset where the positive band orbital magnetization is plotted as a function of the spectral weight. (c) Non-interacting orbital magnetization as a function of chemical potential $\mu$ for several values of $h$. Two regimes are present: The cyan background delimitates the region where bands are mainly linear and white backgrounds where bands have quadratic behavior. The inset is a zoom of the same figure in the linear regime. Marks represent the numerical value from \eref{Eq.orbital_diagram} and dashed lines are the functions $\mu \sigma_{xy}(h)$ with $\sigma_{xy}(h)$ the Hall conductivity in the absence of interaction but with a Zeeman term $h$. (d) Interacting orbital magnetization as a function of the electronic density near half-filling for $h=0.05$. The inset is the same plot as a function of chemical potential.} \label{Fig.orb_tot}
\end{center}
\end{figure*}

\subsection{Orbital magnetization for the non-interacting case}
Similar to the AHE, the orbital magnetization is non-zero only in the presence of a finite $h$. In order to understand how a finite $h$ leads to a finite orbital magnetization, we look at the contribution of each component (partial magnetization) to the trace in \eref{Eq.orbital_diagram}. Panel (a) of Fig.~\ref{Fig.orb_tot} shows the partial orbital magnetizations.  As one can see, the partial orbital magnetizations are non-zero even at $h=0$ but they have opposite signs, canceling each other at $h=0$. At finite $h$, the exact cancellation between the partial orbital magnetizations disappears, leading to finite orbital magnetization.  However, in the model~\eref{EqHamiltonian_total}, the particular form of partial orbital magnetization and the fact that the shift is symmetrical around $\mu=0$ imposes zero orbital magnetization for $\mu=0$, regardless of the value of $h$. In terms of symmetries, even if the Green's function part of the orbital magnetization~\eref{Eq.orbital_diagram} is affected at finite $h$, the particle-hole symmetry \eref{Eq:particle_hole} is still present when $\mu=0$. The presence of the energy vertex and particle-hole symmetry then imply that the orbital magnetization vanishes at $\mu=0$ even with the Zeeman term $h$. These symmetry considerations carry over in the interacting case.  

In an undoped non-interacting topological-insulator the orbital magnetization is proportional to the Chern number when the chemical potential lies within the energy band gap~\cite{Orbital_in_gap,Nourafkan_2014}. Indeed, in 2D, a Chern insulator carries a net dissipationless chiral current on its edge that should contribute to the orbital magnetization\cite{Thonhauser_2005}.
 Our results show that a similar relationship is valid in a non-interacting ferromagnetic Weyl metal as long as the chemical potential lies in the linearly dispersive band structure, i.e, when $-2 \lesssim \mu \lesssim 2 $.  Panel (c) of Fig.~\ref{Fig.orb_tot} displays the orbital magnetization of the non-interacting system as a function of chemical potential for several $h$ values. The cyan background delimitates the region where bands are mainly linear. In this region the  orbital magnetization is proportional to the AHE with the chemical potential as coefficient.  In other words,  for this range of chemical potentials, the following relation holds 
\begin{equation} \label{Eq.orb_WSM}
M^{z}_{\text{orb}} = -\mu \sigma_{xy}.
\end{equation}

Two facts explain this relation. First, for non-interacting systems, it has been shown that, up to fundamental constants, the intrinsic part of the AHE is equal to the derivative of the orbital magnetization with respect to the chemical potential, the St\v{r}eda formula~\cite{Streda_1982,Ito_2017}. Second, even if the Fermi energy does not lie on Weyl nodes, as long as the dispersion is still linear and the Fermi surface can be decomposed in disconnected sheets, the anomalous Hall conductivity reduces to its intrinsic value~\cite{Burkov_2014} defined in \eref{Eq.value_sigdc}, so that it does not depend on chemical potential. Combining these two results, the orbital magnetization should reduce to \eref{Eq.orb_WSM}. This is verified by the linear behavior of the magnetization near the Fermi energy (cf. Panel (c) of Fig.~\ref{Fig.orb_tot}). When the conditions previously enumerated are present, the slope is equal to the anomalous Hall conductivity at half-filling. This argument is not valid for $|\mu| \gtrsim 2$ where non-linear contributions of the band dispersion are not negligible anymore, marking at the same time the end of the linear behavior of the orbital magnetization. 

\subsection{Orbital magnetization for the interacting case}

The effect of interactions on the orbital magnetization is complex. In contrast to the AHE, orbital magnetization is not a quantum integer and therefore it is not protected against perturbations. Algebraically, this can be seen by comparing the integrands in \eref{AHC}, second line, and in \eref{Eq.orbital_diagram}.
In the anomalous Hall conductivity, the frequency vertex, \emph{i.e} the derivative of the inverse Green's function with respect to the Matsubara frequency, plays an important role in making the integrand of the anomalous Hall conductivity interaction-independent of spectral weight at low temperature (see. Appendix \ref{Sec:AppendixB} for a detailled demonstration using the QP approach). This factor is absent in \eref{Eq.orbital_diagram}.
Panel (b) of Fig.~\ref{Fig.orb_tot} shows the positive band partial orbital magnetization of the half-filled case as a function of $U$ in the absence of $h$. Upon increasing $U$, the partial orbital magnetization decreases. This is a consequence of a reduction of the coherent spectral weight, as one can see from the inset of this panel, which shows an almost linear dependence of the positive band partial orbital-magnetization on $Z$. The fact that interaction suppresses the partial orbital magnetization is true for the doped case as well. 

On the other hand, we know that interaction enhances the effective Zeeman coupling and hence, judging from the non-interacting case, it should increase the total orbital magnetization.  

Clearly then, interactions introduce two competing mechanisms.
Panel (d) in Fig.~\ref{Fig.orb_tot} shows the total orbital magnetization as a function of the electronic density in the Weyl semimetal regime. The magnetization is weakly affected by interaction for $U \lesssim 8$ which means that the two opposing trends almost compensate each other. When $U$ becomes of the same order as the bandwidth ($W = 12$), orbital magnetization becomes more and more sensitive to interactions. They change radically its behavior. The inset in panel (d) shows the total orbital magnetization as a function of chemical potential. The total orbital magnetization remains linear in $\mu$ near half filling with a slope that is weakly affected by electronic correlations for small $U$. When $U\approx 12$, deviations start to be visible in the slope of the orbital magnetization.

Because of the presence of the energy vertex and particle-hole symmetry, \eref{Eq:particle_hole}, the orbital magnetization is unaffected by the Zeeman term $h$ at $\mu=0$ for all values of $U$, as in the non-interacting case. 

\begin{figure}
    \begin{center}
    \includegraphics[scale=0.43]{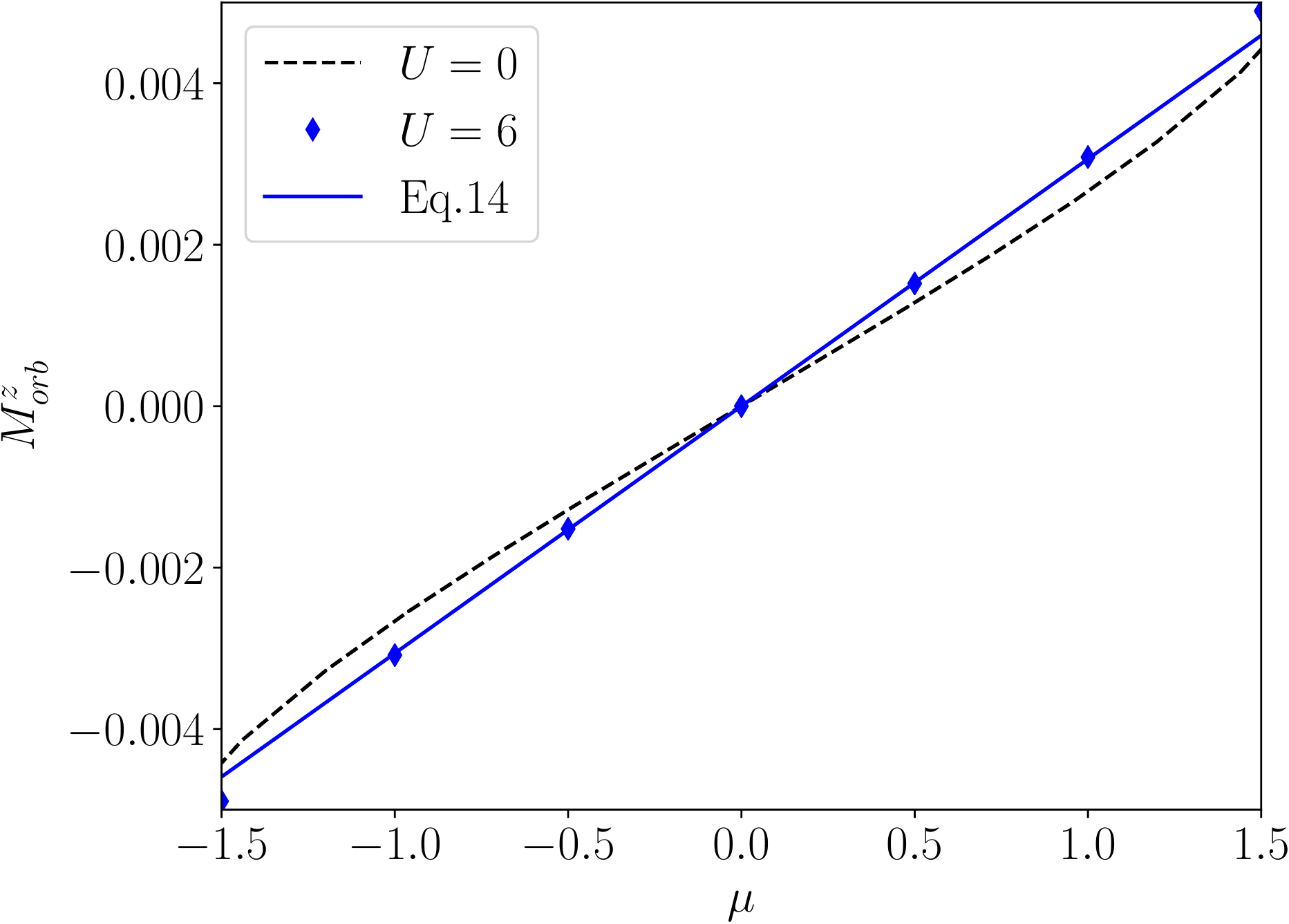}
    \caption{Comparison between numerical results for the the orbital magnetization and the modified St\v{r}eda formula \eref{Eq:ModifiedStreda} where the derivative of $\Sigma^I$ is obtained using a finite difference method on the self-energy computed to second-order in perturbation theory.} \label{Fig.Streda}
    \end{center}
\end{figure}

In order to check whether or not an equation similar to the St\v{r}eda formula \eref{Eq.orb_WSM} remains valid in the interacting case, we expand the energy vertex in \eref{Eq.orbital_diagram} with respect to the chemical potential and keep the first order. Then, using the QP Hamiltonian, we obtain
\begin{equation}\label{Eq:ModifiedStreda}
M^{QP}_{orb} = -\mu Z^2 \left(1-\frac{1}{2}\frac{\partial \Sigma^I}{\partial \mu}\right) \sigma^{QP}_{xy}, 
\end{equation}
where $\sigma^{QP}_{xy}$ is the interacting Hall conductivity obtained from the QP approximation. For small to intermediate values of $U$, the above equation can reproduce quantitatively the change of the orbital magnetization near half-filling, as seen in Fig.~\ref{Fig.Streda}. In the strong correlation regime, however, the orbital magnetization is not linear anymore, aiming towards zero as we approach the Mott transition.

\section{Conclusion}

We have studied a Hubbard model of Weyl semimetals on the cubic lattice~\cite{Roy_2016}, \eref{EqHamiltonian_total} and \eref{Eq.Hamiltonian_nonint}, that breaks time-reversal symmetry and parity and that contains four Weyl nodes (Table I). Even in the presence of $U$, the model is particle-hole symmetric at half-filling, Eq.~(\ref{Eq:particle_hole}), whatever the value of the Zeeman term $h$. The model also has two translated mirror symmetries, Eqs.~(\ref{Eq.symA})(\ref{Eq.symB}), valid for arbitrary interaction and filling when $h=0$. Within the DMFT framework, we have shown that large enough interactions lead to a transition either to a Mott insulator or to a band insulator. Before these transitions, in the semi-metal regime, the anomalous Hall effect and the orbital magnetization are affected very differently by electronic interactions.

Interactions modify the anomalous Hall effect at half-filling by affecting spin polarization when there is a finite Zeeman term $h$. This physics can be explained with the quasiparticle formalism. Interactions change the effective distance between Weyl nodes and the anomalous Hall effect is still controlled by the distance between nodes in momentum space, as in the non-interacting case. So, even in the presence of a Hubbard $U$, the topological properties of this Weyl semimetal, at the origin of the intrinsic anomalous Hall effect, are in a sense protected. When $h=0$, the anomalous Hall effect at half-filling vanishes even in the presence of interactions because the translated mirror symmetries Eqs.~(\ref{Eq.symA}) and (\ref{Eq.symB}) force the effective Weyl node separation to be a reciprocal lattice vector. 

In contrast with the intrinsic part of the anomalous Hall effect, which involves only the kinetic energy or equivalently the renormalized band structure, the orbital magnetization involves both kinetic and potential energies. In the absence of potential energy, the St\v{r}eda formula gives a simple relationship between those two quantities. However, Hall conductivity and orbital magnetization are affected differently by interactions. While a quasiparticle approach is sufficient to accurately understand the change in the anomalous Hall conductivity, it fails for the orbital magnetization because of the potential energy contribution. 

The orbital magnetization at half-filling vanishes for all $U$ and $h$ as a consequence of particle-hole symmetry \eref{Eq:particle_hole}. When $h$ is finite, there is a finite orbital magnetization but it is weakly affected by interaction, even if it is \emph{not} topologically protected. This relative robustness is due to two competing mechanisms: the increase of the anomalous Hall effect due to interaction and the decrease of quasiparticle weight. These effects compensate almost exactly when the chemical potential is not too far from the Weyl nodes. An approach that mixes quasiparticle approach and second order perturbation theory for the self-energy allowed us to find a modified St\v{r}eda formula \eref{Eq:ModifiedStreda} that is accurate for small values of Hubbard interaction. In the strongly correlated regime, the orbital magnetization vanishes due to the interaction-induced localization. 

Our work highlights the unusual properties of orbital magnetization and anomalous Hall effect in the presence of interactions. Compounds like the antiferromagnet $\text{Mn}_3\text{Sn}$ would be interesting to study since the spin magnetization should not dominate over the orbital magnetization. 

\begin{acknowledgments}
We thank Ion Garate, P.L.S Lopes, P. Rinkel and S. Bertrand for useful discussions. This work has been supported by the Natural Sciences and Engineering Research Council of Canada (NSERC) under grant RGPIN-2014-04584,the Canada First Research Excellence Fund and by the Research Chair in the Theory of Quantum Materials. Simulations were performed on computers provided by the Canadian Foundation for Innovation, the Minist\`ere de l'\'Education des Loisirs et du Sport (Qu\'ebec), Calcul Qu\'ebec, and Compute Canada.
\end{acknowledgments}

\appendix
\section{Non-interacting Hamiltonian and effective Hamiltonian calculation}\label{Sec:AppendixA}
The non-interacting Hamiltonian breaks time-reversal symmetry and has two bands (cf. \eref{Eq.Hamiltonian_nonint}):
\begin{eqnarray}
\nonumber
\epsilon_{\pm}= &\pm & \left((h-2t\left(\cos(k_x)+\cos(k_y)+\cos(k_z)\right))^2\right. \\
&+& \left. 4t^2\sin^2(k_x) + 4t^2\sin^2(k_y)\right)^\frac{1}{2}.
\end{eqnarray}
Depending on the value of $h$, the model exhibits a different number of Weyl nodes. Four Weyl nodes are present in the first Brillouin zone when $|h|<2t$ while the number of nodes decreases to two when $2t\leq|h|<6t$. The system is fully gapped for large values of $h$ (\emph{i.e.} $h>6t$). For this paper, we set $h\ll 2t$. The particle-hole symmetry \eref{Eq:particle_hole} imposes $\mu=0$ at half-filling regardess the value of the Zeeman term. \\

In order to find out the anomalous Hall effect \eref{Eq.value_sigdc}, we can use the two low-energy Hamiltonians
\begin{equation}\label{Eq.eff_ham}
H^{\text{eff}}_\pm(q_x,q_y,k_z) = (h-2t\cos(k_z))\boldsymbol{\sigma}_z \pm 2tq_x \boldsymbol{\sigma}_x \mp 2tq_y\boldsymbol{\sigma}_y. 
\end{equation}
These are low-energy models along the $x$ and $y$ directions, but no approximation is done for the $z$ direction. Each Hamiltonian gives two Weyl nodes located at $k_z = \pm \arccos(h/2t)$ and have the same $z$-component of Berry curvature for the \textit{negative} energy band:
\begin{equation}
\Omega^z_{-} =  -\frac{2t^2(h-2t\cos(k_z))}{((h-2t\cos(k_z))^2+4t^2q^2_x+4t^2q^2_y)^{\frac{3}{2}}}.
\end{equation}
The intrinsic anomalous Hall conductivity of one Weyl node at half-filling where only the \textit{negative}-energy band contributes is:
\begin{eqnarray}
\sigma^{\text{eff}}_{xy} &=& \frac{1}{8\pi^3}\int dk_z d^2q \ \Omega^z_-({q_x,q_y,k_z}) 
\\ 
&=&-\frac{1}{8\pi^2}\int^{\pi}_{-\pi}dk_z {\rm sgn} (h-2t\cos(k_z)) \\
&=& -\frac{1}{2\pi^2}\left[\frac{\pi}{2} -\arccos(\frac{h}{2t}) \right] \\
&=& -\frac{1}{2\pi^2}\arcsin(\frac{h}{2t}).
\end{eqnarray}
Here, the function $sgn$ gives the sign of its argument. Since each effective Hamiltonian in ~\eref{Eq.eff_ham} gives the same result, we finally recover \eref{Eq.value_sigdc} of the main text.

\section{Quasiparticle approximation for the lattice model}\label{Sec:AppendixB}

In this appendix, we derive analytic expressions for the calculation of the anomalous Hall conductivity of the lattice model in the quasiparticle approximation.

The quasiparticle Green's function is defined as $\boldsymbol{G}^{QP} = \mathbf{Z}^{1/2}\left[i\omega_n \textbf{I} - \boldsymbol{H}^{QP} \right]^{-1}\mathbf{Z}^{1/2} = Z\tilde{\boldsymbol{G}}$ since the spectral weight matrix is diagonal and has identical values for the two spin species. The previous expression is valid for small value of $i\omega_n$, but since the Fermi-surface physics comes from small values of Matsubara frequencies, we expect to find the correct physical expression by using the QP Green's function regardless of the value of the Matsubara frequency.

In this approximation, the expression for the AHE deduced from \eref{AHC} becomes
\begin{eqnarray}
\nonumber
\sigma^{QP}_{xy} &=& -\text{Im} \frac{1}{\beta} \sum_{\textbf{k},\omega_n} \Tr\left[\boldsymbol{v}_x\boldsymbol{G}^{QP}\frac{\partial \boldsymbol{G}^{QP -1}}{\partial i\omega_n}\boldsymbol{G}^{QP}\boldsymbol{v}_y\boldsymbol{G}^{QP}\right] \\
&=& -\text{Im} \frac{Z^2}{\beta} \sum_{\textbf{k}, \omega_n} \Tr\left[\boldsymbol{v}_x\tilde{\boldsymbol{G}}\tilde{\boldsymbol{G}}\boldsymbol{v}_y\tilde{\boldsymbol{G}}  \right].
\end{eqnarray}
For simplicity, we omitted the indices $\textbf{k}$ and $i\omega_n$ in the above expression and used for the vertex $\partial_{i\omega_n} G^{QP-1} = Z^{-1}$ in the QP approximation. Using the expression for the Green's function and inserting complete sets of states, we find 
\begin{widetext}
\begin{eqnarray}
\sigma_{xy} &=& -\text{Im}\frac{Z^2}{N\beta} \sum_{\textbf{k}, \omega_n} \Tr\left[\boldsymbol{v}_x\tilde{\boldsymbol{G}}\tilde{\boldsymbol{G}}\boldsymbol{v}_y\tilde{\boldsymbol{G}}  \right] \\
\nonumber
&=& -\text{Im}\frac{Z^2}{N\beta} \text{Im}\sum_{\textbf{k},i\omega_n} \sum_{n,m}\left[\frac{v^{n,m}_x v^{m,n}_y}{(i\omega_n - \xi^{QP}_m)^2(i\omega_n -\xi^{QP}_n)}\right] \\
\nonumber
&=& -\text{Im}\frac{Z^2}{N\beta}\sum_{\textbf{k},i\omega_n} \sum_{n,m}\left[\frac{\partial}{\partial \xi^{QP}_m}\left(\frac{v^{n,m}_x v^{m,n}_y}{(i\omega_n - \xi^{QP}_m)(i\omega_n -\xi^{QP}_n)}\right)\right],
\end{eqnarray}
where $v^{n,m}_i$ is $\langle n|\textbf{v}_i|m\rangle$, the matrix element $n,m$ of the bare velocity vertex in the band basis, and $\xi^{QP}_n$ are the eigenenergies of the QP Hamiltonian corresponding to the band $n$. 
The sum over Matsubara frequencies does not present any issues and, after some algebra, one finds
\begin{eqnarray}
\nonumber
\sigma_{xy} = &-& \text{Im}\frac{Z^2}{N}\sum_{\textbf{k}} \sum_{\substack{n,m \\ m\neq n}} \left[ \frac{v^{n,m}_x v^{m,n}_y}{(\xi^{QP}_m-\xi^{QP}_n)^2}\left[n_F(\xi^{QP}_m)-n_F(\xi^{QP}_n) \right] \right. \\
 &-& \left. \frac{v^{n,m}_x v^{m,n}_y}{\xi^{QP}_m-\xi^{QP}_n} \frac{\partial n_F(\xi^{QP}_m)}{\partial \xi^{QP}_m} \right]. \label{Eq.Sigma_final}
\end{eqnarray}
\end{widetext}
At low temperature, the derivative of the Fermi-Dirac distribution can be approximated by a Dirac delta $\delta (\xi^{QP}_n)$, hence it is only sensitive to the Fermi surface. Since we study a semimetal with the chemical potential lying on the node, we can neglect the corresponding contribution at half-filling. Furthermore, dispersion energies are directly proportional to $Z$, leading to an expression \eref{Eq.Sigma_final} that depends on $Z$ only in the Fermi-Dirac distribution since the prefactor $Z^2$ is canceled out by the renormalization of the band energies appearing in denominators. 
At low temperature then, within the QP approximation, it appears that the anomalous Hall conductivity has the same expression as in the non-interacting case with only the replacements $h \rightarrow Z(h+\Sigma^z)$ and $t \rightarrow Zt$.

%

\end{document}